\documentclass[
 amsmath,amssymb,
 aps, twocolumn,
 pra, superscriptaddress] {revtex4-1}

\usepackage     			{graphicx}
\usepackage     			{color}
\usepackage				{upgreek}
\usepackage     			{hyperref}
\usepackage     			{braket}
\usepackage				{amsmath}
\usepackage				{siunitx}
\usepackage     			{float}
\usepackage     			{gensymb}
\usepackage	[normalem]	{ulem}

\definecolor{green}{rgb}{0.45, 0.45, 0}
\definecolor{darkred}{rgb}{0.5, 0, 0}
\definecolor{darkblue}{rgb}{0.1, 0.1, 0.5}
\definecolor{gray}{gray}{0.4}
\definecolor{teal}{rgb}{0, 0.4, 0.4}
\definecolor{purple}{rgb}{0.5, 0, 0.4}
\hypersetup{
    colorlinks,
    linkcolor=gray,
    citecolor=darkred,
    urlcolor=green,
    pdftitle={Switching Rydberg interactions by three orders of magnitude using a terahertz field},
    pdfauthor={K. Wadenpfuhl and C. S. Adams},
}

\begin{document}

\title{Switching Rydberg interactions by three orders of magnitude using a terahertz field}

\author{Karen Wadenpfuhl}
\email{karen.wadenpfuhl@durham.ac.uk}
\affiliation{Department of Physics, Durham University, DH1 3LE, United Kingdom}

\author{Aaron Reinhard}
\affiliation{Department of Physics, Kenyon College, 201 North College Rd., Gambier, Ohio 43022, USA}

\author{Oliver Hughes}
\affiliation{Department of Physics, Durham University, DH1 3LE, United Kingdom}

\author{Lucy Downes}
\affiliation{Department of Physics, Durham University, DH1 3LE, United Kingdom}

\author{Kevin Weatherill}
\affiliation{Department of Physics, Durham University, DH1 3LE, United Kingdom}

\author{C. Stuart Adams}
\email{c.s.adams@durham.ac.uk}
\affiliation{Department of Physics, Durham University, DH1 3LE, United Kingdom}

\date{\today}

\begin{abstract}
Atom-based quantum computing exploits the ability to enhance atom-atom interactions by employing laser excitation to higher-excited Rydberg states. Additional fields that drive transitions between Rydberg states can offer independent control of these atom-atom interactions. However, as microwave (mw) fields only provide access to states with similar principal quantum number $n$, their ability to switch the interactions' strength is limited. Here, we use a pulsed terahertz field to rapidly switch the strength of interactions between Rydberg atoms by three orders of magnitude. We demonstrate interaction switching using photon storage, where the terahertz field induces an interaction induced dephasing of the stored photon. This ability to switch interactions offers advantages for single-qubit readout, state-detection schemes, quantum annealing, and Rydberg quantum optics.
\end{abstract}

\maketitle

The precise control of highly-excited Rydberg atoms has led to fruitful developments in quantum simulation and computing \cite{Saffman2010, Saffman2016, Browaeys2020, Morgado2021, Adams2019}, sensing and metrology \cite{Sedlacek2012, Fan2015, OBrien2009, Downes2020, Holloway2018}, and quantum optics \cite{Pritchard2010, Peyronel2012, Dudin2012, OrnelasHuerta2020, Busche2017, Firstenberg2016, Kumlin2023}. Quantum computing and quantum optics are primarily interested in interactions between Rydberg atoms, whereas quantum sensing relies more heavily on transitions between Rydberg states. To date, most work on interactions has made use of direct dipole-exchange interactions \cite{Ravets2014, Emperauger2025} or has focused on second-order interactions between atoms in the same Rydberg state \cite{Urban2009, Bernien2017, Walker2005, Walker2008}. 

While most work involving transitions in the Rydberg manifold has been restricted to the radiofrequency (rf) or microwave (mw) range ($0.1-60$~GHz) where widely-tunable, high-power sources are readily available, many transitions fall into the higher-energy terahertz (THz) range ($0.1-10$~THz). Terahertz fields allow coupling between Rydberg states with larger change in principal quantum number $\Delta n=\vert n_2-n_1\vert$, which can have dramatically different interaction strengths. The ability to rapidly switch interaction strengths within the Rydberg manifold has advantages for quantum annealing protocols \cite{Glaetzle2017, Angkhanawin2026}, and would be particularly useful in the case of Rydberg quantum optics, where it decouples light propagation from interactions \cite{Paredes2014}. 
However, terahertz frequency fields are not commonly used in Rydberg atomic physics due to the technical difficulties inherent in their creation and detection. Energies of THz photons are above the microwave regime where electronic detectors fail, but are too low for semiconductors that are commonly used for the optical and infrared ranges \cite{Leitenstorfer2023}. Widely-tunable, high-power, on-demand THz source systems comparable to current microwave technology are still in development. Using and probing THz transitions with Rydberg atoms therefore opens new possibilities for sensing and imaging \cite{Downes2020, Chen22, Krokosz25}.

In this work, we demonstrate the use of a terahertz field to switch the strength of Rydberg interactions by three orders of magnitude. A terahertz field with a frequency of 0.6~THz and nanosecond pulse duration coherently drives transitions between Rydberg states. To demonstrate interaction switching, we select states with relatively low principal quantum numbers ($n\sim 35)$ that have interaction strengths differing by three orders of magnitude. The experiments are performed on photons stored as collective excitations, so-called Rydberg polaritons, in an ultra-cold atom ensemble \cite{Busche2016, Busche2017}. We observe strong interaction-induced dephasing of the polaritons when the terahertz field is applied.

\begin{figure}
\centering
\includegraphics[width=\linewidth]{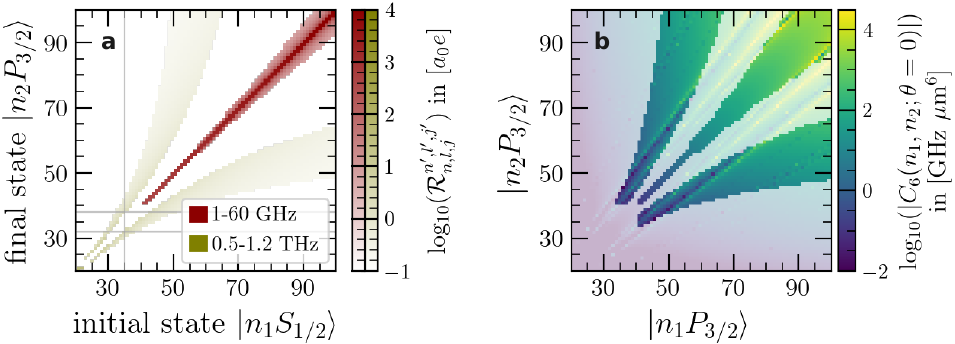}
\caption{\textbf{Typical $\mathbf{\ket{n_1 S_{1/2}} \leftrightarrow \ket{n_2P_{3/2}}}$ transitions within rubidium Rydberg manifold and $\mathbf{C_6(n_1, n_2)}$ interaction map of $\mathbf{P_{3/2}}$ pair states.}
Transitions $\ket{n_1 S_{1/2}} \to \ket{n_2 P_{3/2}}$ within the Rydberg manifold that fall into different frequency bands are shown in \textbf{a}. A source operating in the THz band from 0.5 THz to 1.2 THz gives access to transitions with a large difference in principal quantum number $\Delta n = n_2 - n_1$ even at low $n$. The shading is proportional to the radial dipole matrix element of the transition. Solid lines indicate the states used in this work ($35S_{1/2}$, $32P_{3/2}$, $38P_{3/2}$). \textbf{b} shows the interaction strength $C_6(n_1 P_{3/2}, n_2 P_{3/2};\ \theta=0)$ for rubidium. F\"orster resonances are visible as bright lines in the plot. The central highlighted area of states $\ket{n_1P_{3/2}, n_2P_{3/2}}$ with $n_1 \approx n_2$ can be reached from a $\ket{nS_{1/2}, nS_{1/2}}$ state via two microwave transitions in the range $1 - 60$~GHz. The other two highlighted areas can be reached from $\ket{nS_{1/2}, nS_{1/2}}$ with a mw transition in the $1-60$~GHz range and a THz transition in the range $0.5 - 1.2$~THz.}
\label{fig:Figure1}
\end{figure}

A comparison of mw and THz transitions in rubidium are shown in Figure \ref{fig:Figure1} \textbf{a}. Transitions between an $\ket{nS_{1/2}}$ and an $\ket{nP_{3/2}}$ state that are accessible via a mw field (1-60 GHz, red shading) have small differences in principal quantum number $\Delta n$, whereas transitions coupled via a THz field (0.5 - 1.2 THz) allow access to states with much larger $\Delta n$, as can be seen by the green shaded areas. In Fig.~\ref{fig:Figure1} \textbf{b} we illustrate the pair states that can be reached via two microwave transitions (highlighted region along the diagonal with $n_1 \approx n_2$) and their interaction strength $C_6$. Since terahertz fields can couple to states with larger $\Delta n$ one can address pair states with dramatically different interaction strengths, as indicated particularly by the bright lines in the two outer highlighted areas of Figure \ref{fig:Figure1} \textbf{b}. The THz field therefore allows driving between pair states whose interaction strengths span orders of magnitude, such as we demonstrate in this work. This promises to be useful for Rydberg state-detection schemes and qubit readout sequences \cite{Xu2021, Sumarac2026}, or quantum annealing protocols \cite{Glaetzle2017, Angkhanawin2026}.

\begin{figure}
\centering
\includegraphics[width=\linewidth]{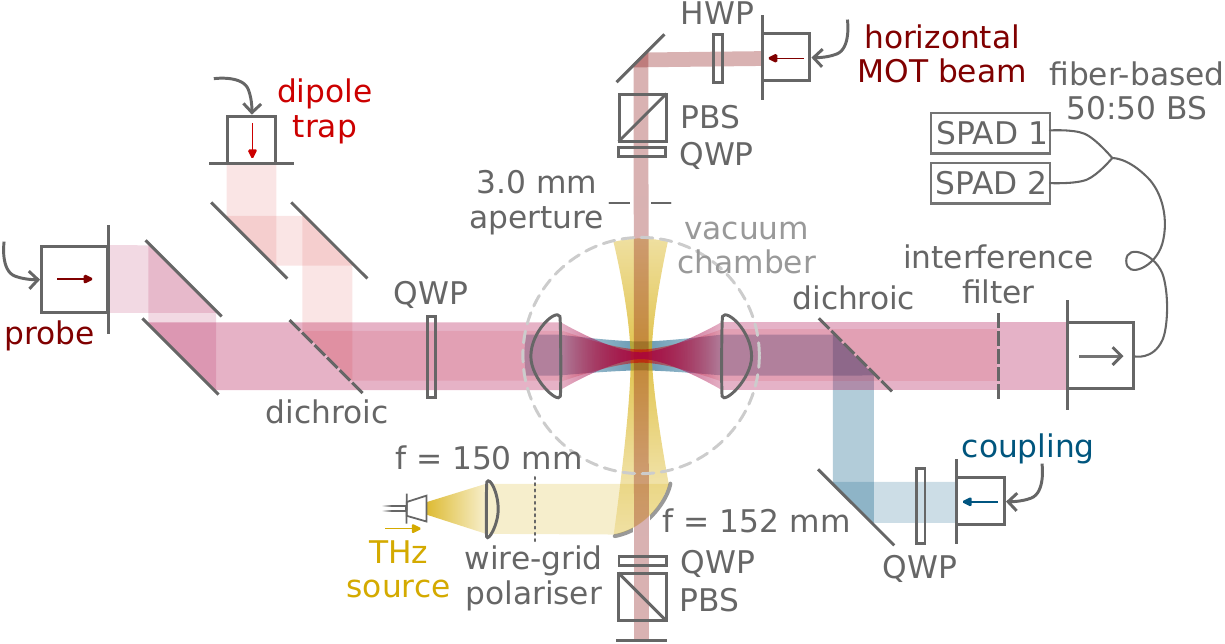}
\caption{\textbf{Simplified schematic of the experiment setup including THz optics (view from above).}
A 3D MOT (horizontal beam shown in dark red, vertical beam pair not shown) cools and traps $^{87}$Rb atoms in the center of the vacuum chamber before they are loaded into an optical dipole trap (orange). Counterpropagating probe (red) and coupling lasers (blue) excite the atoms to a Rydberg state. Photons on the probe transition are detected with a SPAD pair, the dipole trap light is blocked from the SPADs with interference filters. A THz source output (gold) is collimated by a teflon lens, polarisation-cleaned with a wire-grid polariser, and focused onto the atoms with a parabolic mirror. The parabolic mirror has a hole in the center with 3.0 mm clear aperture to which the horizontal MOT beam is size-matched.
}
\label{fig:Figure2}
\end{figure}

\begin{figure}
\centering
\includegraphics[width=\linewidth]{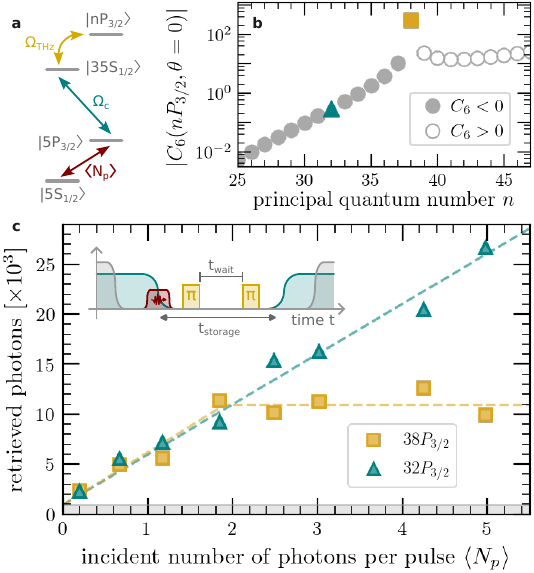}
\caption{\textbf{Saturation of photon storage in $\mathbf{\ket{38P_{3/2}}}$ for high incident photon numbers.}
Probe (red) and coupling (blue) lasers address the $\ket{5S_{1/2}, F=2} \to \ket{5P_{3/2}, F^\prime=3} \to \ket{35S_{1/2}}$ transitions, as shown in the level scheme in \textbf{a}. The THz is used to coherently drive the excitations in the Rydberg manifold $\ket{35S_{1/2}} \to \ket{nP_{3/2}}$ with $n = 32,\ 38$ in this work. \textbf{b} shows the interaction strength $C_6(nP_{3/2})$ at orientation $\theta=0$ relative to the quantisation axis. The states addressed in this work are $n = 32,\ 38$, as indicated by triangle and square respectively. \textbf{c} shows the number of retrieved photons as a function of incident probe photon number per pulse, $\braket{N_p}$. A linear relationship of incident and retrieved photon numbers is found for THz driving to the $\ket{32P_{3/2}}$ state. For the $\ket{38P_{3/2}}$ state, the initial increase saturates at $\braket{N_p} \approx 2$ incident photons per pulse. Dashed lines are guides to the eye to highlight the different trends. The inset schematically shows the pulse sequence with the photon colours matching that in \textbf{a} and gray representing the dipole trap. The storage duration on the $\ket{nP_{3/2}}$ state is $t_\mathrm{wait} = 100$~ns and the overall photon storage time is $t_\mathrm{storage} = 700$~ns.
}
\label{fig:Figure3}
\end{figure}

The experimental setup is illustrated in Fig.~\ref{fig:Figure2}. $^{87}$Rb atoms from a 2D MOT located above the science chamber are loaded into a 3D MOT (dark red) formed in the center between the in-vacuo lens pair. A subsequent molasses stage cools the atoms to $T = 14(2)$~$\upmu$K before they are loaded into a dipole trap (orange) with typical axial trap frequency $\omega_{ax} \approx 2\pi \times 1.5$~kHz and trap ellipticity $l_{ax}/l_{rad} \approx 23$. Counterpropagating probe (red) and coupling (blue) lasers address the $\ket{5S_{1/2}, F=2} \to \ket{5P_{3/2}, F^\prime=3}$ transition at 780.2~nm and $\ket{5P_{3/2}, F^\prime=3} \to \ket{35S_{1/2}}$ at 480.2~nm, respectively.  Photons on the probe transition are detected with a single-photon avalanche detector (SPAD) pair and cumulative counts from multiple repetitons of the experiment are taken as datapoints. Probe photons are first stored in the ensemble and then retrieved by switching the coupling laser first off and then back on \cite{Fleischhauer2000, Liu2001, Phillips2001}, as shown in the inset of Figure \ref{fig:Figure3} \textbf{c}. During storage we apply the terahertz field to drive the Rydberg excitation to a different state in the Rydberg manifold, see Fig.~\ref{fig:Figure3} \textbf{a} for the level scheme.

The THz radiation is emitted by an amplifier multiplier chain (AMC; WR380 (WM1.5) manufactured by Virginia Diodes with an output frequency range of 0.5-0.75~THz) that multiplies and amplifies the frequency of an input microwave seed signal by a factor of $\times 54$. This AMC and the THz optics are ex-vacuo devices. The AMC output is shaped with a diagonal horn antenna for the appropriate frequency range, collimated by a 2" teflon lens and polarisation-cleaned with a wire-grid polariser. It is then coupled into the horizontal MOT beam path via a parabolic mirror with a 3.0~mm diameter hole in the center, and the horizontal MOT beam waist is matched to this aperture. The parabolic mirror serves as the combining element for the beam path, as well as the final focusing optics for the THz radiation. We estimate a THz beam waist of $w_{0} = 1.2$~mm in the atom plane and a Rayleigh range of $z_R \approx 9$~mm.

A 3D-printed cage adapter plate is mounted on the AMC, and the THz optics, including the parabolic mirror, are mounted in a cage system to simplify alignment. Initial positioning of the collimation lens with a ruler proved to be a good starting point. Careful alignment of the cage system by eye proved sufficient to get an initial overlap of the atom cloud and the THz focus. The collimator lens position and parabolic mirror alignment are then optimised with the atom signal to show maximum Autler-Townes (AT) splitting of a THz-dressed EIT feature. The major THz power loss factor from source to atoms is our vacuum viewport, we have measured power losses of $\sim 50\%$ on a similar vacuum viewport with a THz calorimeter.

The AMC is seeded from a microwave source such that the frequency of the THz radiation is controlled via the microwave seed input frequency. The THz output power can be controlled via an internal attenuation unit of the AMC and the effective THz power at the atoms can be measured by determining the AT splitting of a THz-dressed EIT feature in a resonant 4-level scheme, or by measuring the Rabi frequency on resonance of the THz transition. An electric field quantisation axis is applied along the probe direction, i.e. along the atom cloud's axial direction, and the THz field is polarised along this axis to drive $\pi$ transitions.

Pulsing of the THz output is achieved by pulsing the microwave seed, rather than turning the AMC itself on/off. Switching of the AMC leads to rise and fall times on the order of milliseconds whereas switching of the microwave seed input leads to a THz output with a rise time comparable to that of the seed. By pulsing the microwave seed on ns timescales, we can therefore generate THz pulses of nanosecond duration. Experimentally, we find that when switching on the mw source, the duration of the first $\pi$-pulse is increased by $\sim$10~ns. This behaviour is observed independently of the Rabi frequency of the transition and is attributed to the effective rise time of the THz output, which is a combination of the rise time of the microwave source ($20\% - 80\%$ switching duration of 10 ns) and the AMC.

To demonstrate the ability of the AMC to create coherent, nanosecond pulses in the THz range of the spectrum, we manipulate the Rydberg state of the stored excitation by addressing two transitions from $\ket{35S_{1/2}}$ that fall into the THz band: $\ket{35S_\text{1/2}} \to \ket{32P_{3/2}}$ with $f_{32} = 577.7$~GHz and $\ket{35S_{1/2}} \to \ket{38P_{3/2}}$ with $f_{38} = 607.8$~GHz. Both transitions have large dipole moments $\sim\mathcal{O}$(10 debye) and for each transition the THz source output is attenuated to match a 50~ns duration for the first $\pi$-pulse, which includes the $\sim 10$~ns 20\%-80\% rise time of the microwave seed. 

Figure \ref{fig:Figure3} \textbf{b} shows the interaction strength $C_6(nP_{3/2})$ with principal quantum number $n$ for the range of interest. The $\ket{38P_{3/2}}$ state (yellow square) is nearly degenerate with a dipole-coupled pair-state, a so-called F\"orster resonance, that causes a strong second-order interaction $C_6(38P_{3/2}) = -307.5$~GHz~$\upmu$m$^6$. Since $C_6(32P_{3/2}) = -0.29$~GHz~$\upmu$m$^6$ (teal triangle) is much lower, we can address states with interaction strengths differing by three orders of magnitude. The interaction strength of the initial storage state $\ket{35S_{1/2}}$ is $C_6(35S_{1/2}) = -0.19$~GHz~$\upmu$m$^6$. All $C_6$ values are stated for $m_\mathrm{j_1} = m_\mathrm{j_2} = \pm 1/2$ since we drive $\pi$-transitions with the THz field.

Figure \ref{fig:Figure3} \textbf{c} shows the retrieved photon counts after $t_\mathrm{wait} = 100$~ns (interaction time in the $\ket{nP_{3/2}}$ state) as a function of incident probe photon number $\braket{N_p}$, the pulse sequence is shown in the inset. For the $\ket{32P_{3/2}}$ state (triangles) we observe a linear increase in retrieved photon number as $\braket{N_p}$ increases. The retrieved photon counts for storage in $\ket{38P_{3/2}}$ (squares) saturates at $\braket{N_p} \approx 2$, which is attributed to interaction-induced dephasing of multiple excitations (see Supp. Material \ref{app:levellingOffModel}).

\begin{figure}
\centering
\includegraphics[width=\linewidth]{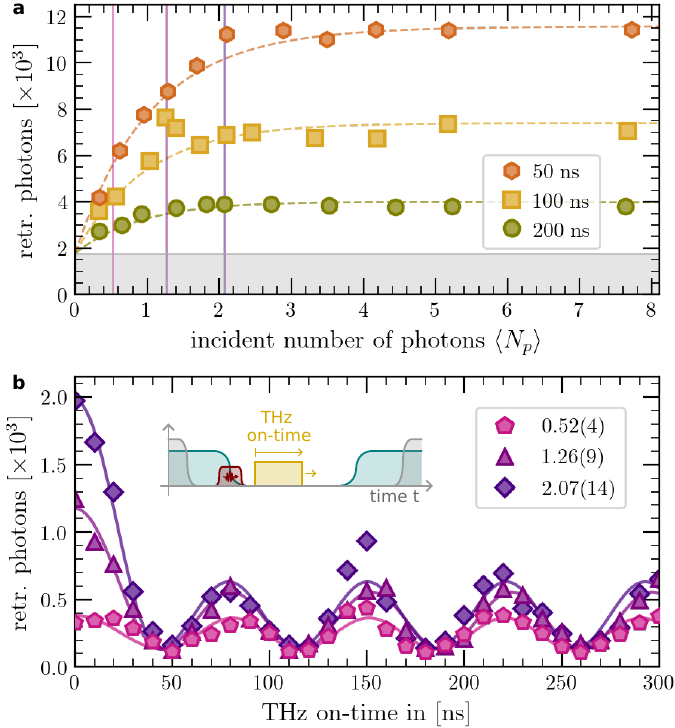}
\caption{\textbf{Interaction-induced suppression of Rabi oscillation amplitude at $\mathbf{\ket{38P_{3/2}}}$.}
The retrieved photon number as a function of incident probe photon number $\braket{N_p}$ is shown in \textbf{a} for different wait times $t_\mathrm{wait}$ in the $\ket{38P_{3/2}}$ state. The overall storage time $t_\mathrm{storage} = 700$~ns is the same for all three datasets. A Monte-Carlo-based dephasing model (see Supp. Material \ref{app:levellingOffModel}) is fitted to the data and shown as dashed lines. The background count level is indicated in gray. In \textbf{b}, the photon number retrieved from $\ket{35S_{1/2}}$ is shown for a single THz pulse of varying duration, as indicated in the inset. The resulting Rabi oscillations on the $\ket{35S_{1/2}} \leftrightarrow \ket{38P_{3/2}}$ transition show a suppression in amplitude after an initial cycle through the $\ket{38P_{3/2}}$ state for incident photon numbers $\braket{N_p} =$ 1.26(9), 2.07(14) (squares and triangles, respectively) but not for $\braket{N_p} =$ 0.52(4) (pentagons). Solid lines show a phenomenological fit to the data, details are given in Supp. Material \ref{app:suppressedRabiOscillations}.
}
\label{fig:Figure4}
\end{figure}

This saturation of retrieved photon number in the $\ket{38P_{3/2}}$ state is studied for different wait times $t_\mathrm{wait}$ between the THz $\pi$-pulse pair, as shown in Figure \ref{fig:Figure4} \textbf{a}. All three show a similar trend for reaching the saturation of retrieval counts at $\braket{N_p} \approx 2$ but have different saturation levels since the retrieval efficiency from the \emph{P}-state varies with wait time, $\eta(t_\mathrm{wait})$, as outlined in Supp. Material \ref{app:levellingOffModel}.

Rabi oscillations on the $\ket{35S_{1/2}} \leftrightarrow \ket{38P_{3/2}}$ transition for different incident probe photon numbers $\braket{N_p}$ are shown in Figure \ref{fig:Figure4} \textbf{b}. For $\braket{N_p} \approx 0.5$ incident photons per pulse, the probability of having more than a single photon per pulse is below $10\%$ and we therefore do not observe significant damping of the oscillation amplitude (pink pentagons). This data demonstrates coherent Rabi oscillations on a transition in the THz range. However, for higher numbers of incident photons per pulse $\braket{N_p} > 1$, the amplitude of the retrieval signal is damped after the first oscillation through the strongly interacting $\ket{38P_{3/2}}$ state. Due to the small ratio of our axial cloud waist $w_{ax}$ to the blockade radius, the probability of more than a single excitation remaining after one Rabi cycle to the $\ket{38P_{3/2}}$ state is very small. We therefore do not see a further damping of the amplitude after the first cycle to the strongly interacting state. The suppression in retrieval amplitude after an initial oscillation through the $\ket{38P_{3/2}}$ state corresponds to an interaction-induced cleanup of the number of stored photons in the polaritonic mode. This can be used to retrieve at maximum only a single excitation, irrespective of the number fluctuations of the Poissonian input field, if the trap geometry is set appropriately relative to interaction strength.

We have demonstrated coherent control of a Rydberg polariton on a transition in the THz range with nanosecond pulse durations. This corresponds to a collectively encoded Rydberg qubit with the qubit states driven on a THz transition \cite{Spong2021}. Due to the challenges associated with the THz range of the spectrum, THz fields have not yet been used for pulsed, coherent control and manipulation of Rydberg qubits before. These challenges range from difficulties in the creation and detection of THz radiation, particularly when pulsed, to challenges in combining the THz with optical beams in a setup. We have shown that it is feasible to build a setup and apply coherent THz pulses to an atomic cloud. 

The THz source unlocks a wide range of transitions within the Rydberg manifold that are inaccessible to microwave fields and makes the F\"orster resonance at $\ket{38P_{3/2}}$ easily accessible from an \emph{S}-state. A combination of THz and mw transition on an excitation pair in a common initial state $\ket{nS_{1/2}, nS_{1/2}}$, e.g. single excitations in two separate traps, gives access to a wide range of pair states $\ket{n_1P_{J_1}, n_2 P_{J_2}}$ with large differences in principal quantum number $\Delta n = n_2 - n_1$. This opens up the possibility to utilise the rich internal structure of Rydberg-atom interaction potentials and, in particular, to address the F\"orster resonances with $n_1 \neq n_2$ at low principal quantum numbers highlighted in Figure \ref{fig:Figure1} \textbf{b}. These strong interactions at low principal quantum numbers $n$ open new possibilities for single-qubit readout in Rydberg atom-based quantum computing and state detection schemes, or for Rydberg atom-based sensing protocols. In particular the availability of strong interactions at low principal quantum numbers is beneficial for use cases where external fields may easily lead to perturbations of the Rydberg state since the sensitivity to external perturbations increases drastically with $n$. Also, larger dipole moments for the optical transition at lower $n$ require less laser power, which is an advantage in particular when taking an application forward to a product.

Additionally, the large dipole moment on the THz transitions used in this work mean that the atoms are very sensitive to the THz electric field and we estimate that our THz $\pi$-pulse contains $\sim\mathcal{O}(10^9)$ THz photons. Other transitions with a higher dipole moment are even more sensitive to the number of incident THz photons, making these transitions potential candidates as sensitive detectors of pulsed THz radiation.

\begin{acknowledgments}
Financial support was provided by the UKRI, EPSRC grant reference number EP/V030280/1, EP/W033054/1 and EP/Z533166/1. AR acknowledges support as a Leverhulme Visiting Professor.

\end{acknowledgments}


\bibliography{bibliography.bib}


\pagebreak
\ \\
\newpage

\onecolumngrid
\appendix

\section*{Supplemental Material for \\ Emergence of synchronisation in a driven-dissipative hot Rydberg vapor}

\section{Monte-Carlo simulation of outgoing photon number with interactions}
\label{app:levellingOffModel}

The retrieval efficiency $\eta$ of a photon stored in a spin wave for time $t$ is given by $\eta = \eta_0 |\braket{\Psi_{t=0} | \Psi_t}|^2$ with $\eta_0$ accounting for experimental limits of the storage and detection efficiency \cite{Mewes2005, SchmidtEberle2020}. Thermal motion of the atoms leads to a loss of retrieval efficiency $\eta(t_{s}) \propto \exp(-t_{s}^2 / \tau_{s}^2)$ with characteristic time $\tau_s = (|\textbf{k}_0| \ v)^{-1/2}$ and average velocity $v = (k_B T / m )^{1/2}$ \cite{Zhao2008, SchmidtEberle2020}. This corresponds to a motional scrambling of the phase of the spin wave that is destructive to retrieval efficiency, unless corrected for with specific rephasing protocols \cite{Viola1998, Rui2015, Jiao2025}, prevented by pinning the atoms in lattices \cite{Dudin2013} or choosing zero-$\textbf{k}$ excitation schemes \cite{Liu2001, Phillips2001, Sibalic2016}.

An additional loss contribution of the retrieval efficiency is introduced by coupling to a second Rydberg state with the Thz source. Again, thermal motion leads to a contribution with characteristic time $\tau_\mathrm{wait} = 1/\sqrt{|\textbf{k}_{Thz}| \ v}$. Additional incoherences in the THz driving and further effects are included and lead to an overall efficiency $\eta(t_\mathrm{storage}, t_\mathrm{wait}) \propto \eta(t_\mathrm{wait}) \exp(-t_\mathrm{storage}^2 / \tau_\mathrm{s}^2)$.

For the experiments presented in the main paper (Figure 3 \textbf{c} and Figure 4) the overall storage time $t_s$ is held constant between different datasets. The general motional loss contribution $\eta(t_\mathrm{storage})$ is therefore constant across different runs and can be absorbed in $\eta_0$.

The loss mechanisms described above are induced by thermal motion and other incoherent processes that lead to dephasing of the spin waves irrespective of the number of excitations stored in the atomic cloud. For strong interactions between Rydberg states, such as the $\ket{38P_{3/2}}$ state, one has to take interaction-induced dephasing between two or more excitations into account as well.

For a poissonian input field with mean photon number $\bar{n}$ and weights $P_m(\bar{n}) = \frac{\bar{n}^m e^{-\bar{n}}}{m!}$, one initially excites the state \cite{Bariani2012}
\begin{equation}
\label{appeqn:PoissonianSpinWave}
    \ket{\Psi^{\bar{n}}_{t=0}} = \displaystyle\sum_{m=0}^\infty \frac{P_m(\bar{n})}{\sqrt{m!}} \left( \hat{S}^\dagger_{\textbf{k}_0} \right)^m \ket{G}
\end{equation}
with the bosonic annihilation operator $\hat{S}_{\textbf{k}_0}$ defined as
\begin{equation}
\label{appeqn:spinWaveAnnihilationoperator}
    \hat{S}_{\textbf{k}_0} = \frac{1}{\sqrt{\mathcal{N}}}  \displaystyle\sum_{\mu=1}^\mathcal{N} \exp(-i\textbf{k}_0 \cdot \textbf{r}_\mu) \ket{g_\mu}\bra{r_\mu}.
\end{equation}
The wave vector $\textbf{k}_0 = \sum_i \textbf{k}_i$ is defined as the effective wave vector of all light fields used during the initial write process of the spin wave.

\begin{figure*}
\centering
\includegraphics[width=\textwidth]{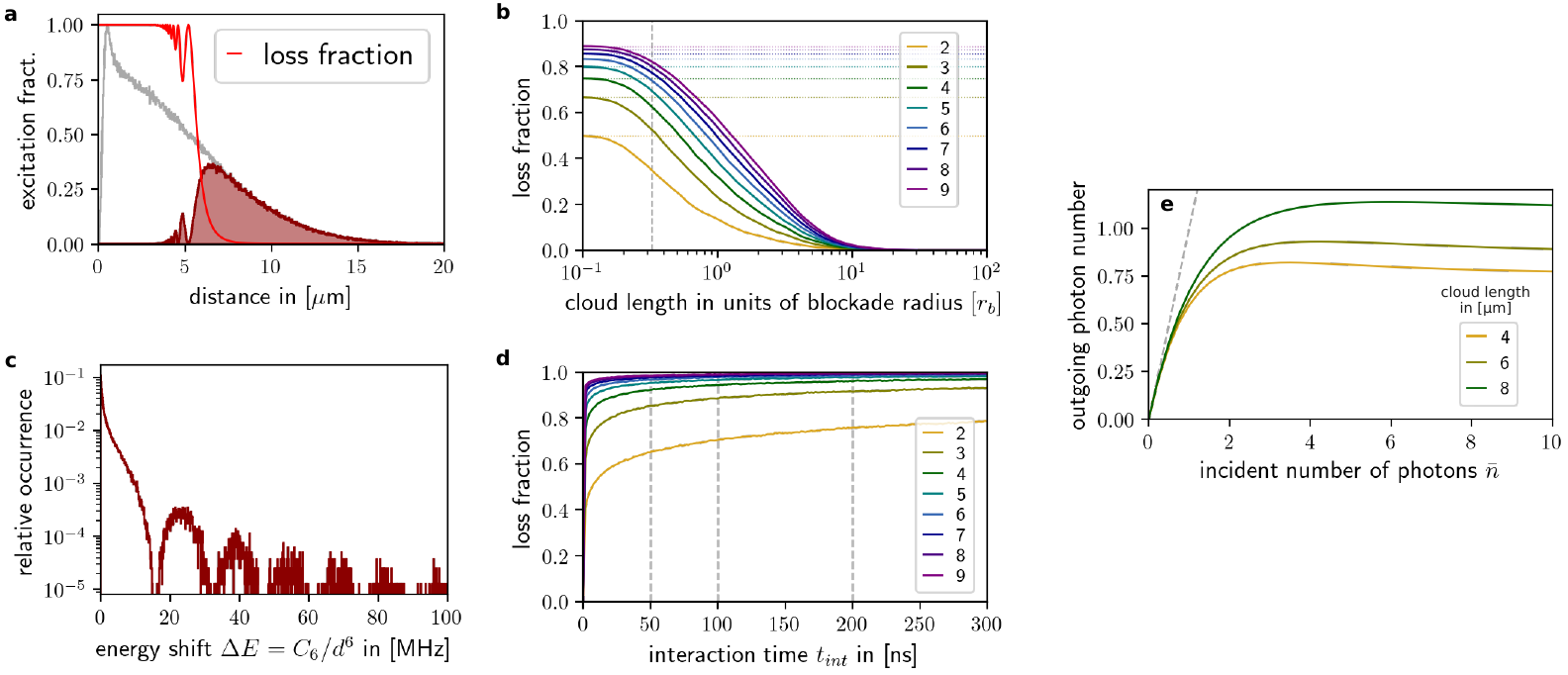}
\caption{\textbf{MC simulation of interaction-induced losses from the polaritonic mode.}
Monte-Carlo simulation of the loss fraction from the polaritonic mode from different loss contributions are shown in \textbf{a} - \textbf{d}. The MC sampled loss fraction during excitation from $\ket{35S_{1/2}} \to \ket{38P_{3/2}}$ is shown in \textbf{a} for our trap geometry. For $m$ excitations, the resulting fractional losses during excitation are shown as a function of axial cloud waist in \textbf{b}, suppressed excitations are treated as being scattered out of the polaritonic mode. In \textbf{c}, the distribution of interaction-induced energy shifts is shown which translates into the interaction-induced dephasing loss fraction shown in \textbf{d}. In \textbf{e}, the resulting outgoing photon number after the second THz $\pi$-pulse with 100~ns wait time is shown as a function of incident photon number $\bar{n}$ for different axial cloud lengths with ellipticity $l_{ax} / l_{rad} = 23$. The gray dashed slope indicates the output photon number without interaction-induced effects and with efficiency $\eta = 1$. For this model curve, we set all other dephasing mechanisms, such as motion-induced dephasing, to zero.}
\label{fig:FigureA1}
\end{figure*}

Interaction-induced blockade and dephasing occur for all instances with more than a single excitation in the trap \cite{Petrosyan2011, Bariani2012, Murray2016, Tian2018}. For $m$ excitations, the interaction-induced phase shift on excitation $j$ follows as $\Delta \phi_j = t_\mathrm{wait}/\hbar \cdot \sum_{i\neq j}^m C_6/r_{ij}^6$ \cite{Bariani2012}. Monte-Carlo sampling of the position distribution for $m$ excitations within a given trap geometry $\sigma_{rad},\ \sigma_{ax}$ allows extraction of the excitation fraction $f_{exc.}(C_6, r_b, \Omega_{Rabi})$ (see Fig. \ref{fig:FigureA1} \textbf{a} and \textbf{b}) \cite{Stanojevic2012} as well as the fractional retrieval efficiency originating from interaction-induced dephasing $\Delta \phi(t_\mathrm{wait}, C_6)$ (see Fig. \ref{fig:FigureA1} \textbf{d}). With these two parameters, one can reconstruct the fractional retrieval losses for $m$ excitations in a given trap geometry and with storage time $t_\mathrm{wait}$. Translating this into output photon numbers after an interaction time $t_\mathrm{wait}$ yields the model curves in Figure \ref{fig:FigureA1} \textbf{e}.

This model is fitted to the data shown in Figure \ref{fig:Figure4} \textbf{a}, with the saturation level left to float since this scales with $\eta(t_\mathrm{wait})$.

\section{Fitting the suppressed Rabi oscillations}
\label{app:suppressedRabiOscillations}

We fit a function of the form
\begin{equation}
\label{appeqn:suppressedRabiOscillationFunction}
    N(t) = 
    \begin{cases}
         & N_0 \cos^2 \left(\pi \frac{t}{2t_0}\right) \text{  if } t \leq t_0\\
         & N_1 \cos^2 \left(\frac{\pi}{2} + \frac{1}{2} \Omega (t-t_0) \right)\text{ if } t > t_0
    \end{cases}
\end{equation}
with Rabi frequency $\Omega$ and the duration of the first $\pi$-pulse $t_0$ a free parameter. The initial $\pi$-pulse duration is longer than what one expects from the Rabi frequency on the transition $\Omega_{THz} = 2\pi \times 14.0(2)$~MHz, which is caused by the combined finite rise time of the microwave source and the AMC. 

The different amplitudes for $t \leq t_0$ and $t > t_0$ account for interaction-induced losses of excitations due to the strong interaction in the $\ket{38P_{3/2}}$ state. This is another manifestation of the saturation of retrieved photon counts for more than a single incident photon, $\braket{N_p} > 1$.

\end{document}